\documentclass[twocolumn,showpacs,amsmath,amssymb,prd]{revtex4}
          
\usepackage{graphicx}
\usepackage{dcolumn}
\usepackage{bm}
         
\newcommand{\be}{\begin{equation}}
\newcommand{\ee}{\end{equation}}
\newcommand{\ba}{\begin{eqnarray}}
\newcommand{\ea}{\end{eqnarray}}

\def\g{\gamma}

\begin{document}
\input{epsf}

\title{High Energy Neutrinos from Novae in Symbiotic Binaries: The
Case of V407 Cygni}

\author{Soebur Razzaque,$^{1}$\footnote{Resident at the Naval Research
Laboratory, Washington, DC 20375, USA; E-mail:
srazzaque@ssd5.nrl.navy.mil} Pierre Jean,$^{2}$ and Olga Mena$^{3}$}

\affiliation{$^1$National Research Council Research Associate,
National Academy of Sciences, Washington, DC 20001, USA}

\affiliation{$^2$Centre d'Etude Spatiale des Rayonnements, CNRS/UPS,
BP 44346, F-31028 Toulouse Cedex 4, France}

\affiliation{$^3$Instituto de Física Corpuscular, IFIC, CSIC and
Universidad de Valencia, Spain}

\begin{abstract} 
Detection of high-energy ($\gtrsim 100$~MeV) $\gamma$ rays by the {\it
Fermi} Large Area Telescope (LAT) from a nova in the symbiotic binary
system V407 Cygni has opened possibility of high-energy neutrino
detection from this type of sources.  Thermonuclear explosion on the
white dwarf surface sets off a nova shell in motion that expands and
slows down in a dense surrounding medium provided by the red giant
companion.  Particles are accelerated in the shocks of the shell, and
interact with surrounding medium to produce observed $\gamma$ rays.
We show that proton-proton interaction, which is most likely
responsible for producing $\gamma$ rays via neutral pion decay,
produces $\gtrsim 0.1$~GeV neutrinos that can be detected by the
current and future experiments at $\gtrsim 10$ GeV.
\end{abstract}


\date{\today}
\maketitle

\section{Introduction}
High-energy ($\gtrsim 1$~GeV) neutrinos are produced dominantly via
decays of charged pions and kaons created by proton-proton ($pp$) and
proton-photon ($p\gamma$) interactions.  Powerful astrophysical
sources of $\gamma$-rays, long duration ($\gtrsim 2$ s) gamma ray
bursts (GRBs); short duration ($\lesssim 2$ s) GRBs and active
galactic nuclei (AGNs), as well as sources such as core-collapse
supernovae, supernova remnants and microquasars have been proposed as
the sources of high-energy $\nu$'s (see e.g. \cite{nu_astro}).
Protons and ions accelerated by a Fermi mechanism in the shocks of
these sources interact with ambient particles and/or soft photons to
produce $\gamma$ rays, via neutral pion decay, and $\nu$'s.  Modeling
$\g$ ray emission from an astrophysical source by a $\pi^0$ model thus
inevitably predicts high-energy $\nu$ flux from the same source (see
e.g. \cite{pi0_models}).  Detection of these $\nu$'s can provide a
conclusive proof of the $\pi^0$ model and discriminate against a
leptonic model for observed $\g$ rays.

Detection of $\gtrsim 100$ MeV $\g$ rays by the {\em Fermi} Large Area
Telescope (LAT) from the Nova 2010 on March 10 in the symbiotic binary
V407 Cygni is the first from any nova \cite{FermiV407Cyg}.  The binary
system consists of a Mira-type pulsating star, a red giant (RG) which
may be swollen to a radius $\sim 500R_\odot$, and a white dwarf (WD)
with mass $M_{\rm WD} \gtrsim 1\,M_\odot$ \cite{Munari90}.  The WD
accretes material from the stellar wind of the RG and forms an
envelope on its surface.  As the mass of the envelope increases, an
increasing WD surface temperature ignites a thermonuclear runaway (see
e.g. \cite{Starrfield89}).  A shell of mass $M_{\rm ej}$ can be
ejected from the WD surface when the proper pressure at the
core-envelope interface exceeds $\sim 2\cdot 10^{20} (M_{\rm
ej}/10^{-6}M_\odot) (M_{\rm WD}/M_\odot) (R_{\rm WD}/10^8\,{\rm
cm})^{-4}$ dyn cm$^{-2}$.  Note, however, that the WD mass and radius
are related roughly as $R_{\rm WD}\propto M_{\rm WD}^{-1/3}$.  Thus a
smaller envelope mass and corresponding $M_{\rm ej}$ is required to
produce a thermonuclear explosion on the WD surface as $M_{\rm WD}$
increases.

An optical nova is detected when the initially hot shell expands to a
large radius from the WD, thus cooled to a temperature $\lesssim 10^4$
K, and when the opacity dropped due to recombination of ionized
hydrogen gas \cite{Starrfield89}.  Non-thermal $\g$ rays detected by
LAT for 15 days following the optical detection requires particle
acceleration and their interactions in the nova shell and in the
surrounding medium of the RG.  Particle acceleration in the shock of
the nova shell has been perceived for the 2006 nova outburst in
another symbiotic binary RS Ophiuchi \cite{Tatischeff07}, but LAT
detection of $\gtrsim 100$ MeV $\g$ rays from V407 Cygni prove
existence of these high-energy particles in the nova shock and their
interactions.

In this brief report we calculate the expected $\nu$ flux from the
nova 2010 in the symbiotic binary V407 Cygni following the $\pi^0$
decay model of $\g$ ray emission from $pp$ interactions reported in
Ref.\ \cite{FermiV407Cyg}.  We also estimate expected $\nu$ events in
current and future detectors and discuss detection prospects of these
type of novae.

\section{Binary system and particle acceleration in nova shell}
Modeling of $\gamma$-ray light curve from the nova 2010 in V407 Cygni
system suggests a binary separation of $a\approx 10^{14}a_{14}$~cm, a
mass-loss rate of ${\dot M}_{\rm w} \approx 3\cdot 10^{-7} A_{\star}
M_\odot$~yr$^{-1}$ from the RG that blows a wind with a measured
velocity of $v_{\rm w} \approx 10 v_{{\rm w,}6}$~km~s$^{-1}$ and a
shell kinetic energy of $E_{\rm k} \approx 10^{44}$~erg with a
measured velocity $v_{\rm ej} = 3200\pm 345$~km~s$^{-1}$
\cite{FermiV407Cyg}.  Close proximity between the WD and RG allows the
parts of the nova shell that expand toward the RG to decelerate and
enter the Sedov-Taylor phase rapidly, resulting in efficient
conversion of shell kinetic energy into shock-accelerated particle
energy.

For a distance $R$ and polar angle $\theta$ from the WD center towards
the RG, the density of particles in the RG wind can be calculated as
$n(R,\theta) = ({\dot M}_{\rm w}/v_{\rm w} {\bar m})(R^2 + a^2
-2aR\cos\theta)^{-1}$ using an iverse square-law density profile of the RG
wind.  The magnetic field in the forward shock that propagates into
the surrounding medium with a velocity $v_{\rm sh} \sim v_{\rm ej}
\approx 3.2\cdot 10^8 v_{{\rm sh,}8.5}$~cm~s$^{-1}$, before
significant deceleration, can be calculated from the thermal energy
density in the RG wind with temperature $T_{\rm w} \approx 700 T_{{\rm
w},2.8}$~K \cite{Munari90} as
\ba
B (R,\theta) = \alpha_B \sqrt{\frac{8kT_{\rm w} 
{\dot M}_{\rm w}/v_{\rm w} {\bar m}}{R^2 + a^2 -2aR\cos\theta}}
\nonumber \\ \approx 0.04
\frac{\alpha_B T_{{\rm w},2.8}^{1/2} A_{\star}^{1/2}
v_{{\rm w,}6}^{-1/2}} {a_{14} - v_{{\rm sh,}8.5} t_{\rm d}}
~{\rm G}; ~\theta =0,
\label{B_field}
\ea
in $t=t_{\rm d}$~day time scale. Here ${\bar m}=10^{-24}$~g is the
mean particle mass; $\alpha_B$ is a magnetic field amplification
factor, which may arise from shock modification, and we assumed a
factor 4 increase in particle density expected to arise in strong
shocks.  The maximum proton energy, accelerated by a Fermi mechanism,
in this magnetic field can be estimated by assuming that the
acceleration time scale is equal to the Larmor time scale as
\ba
E_{p,\rm max} = \frac{eBtv_{\rm sh}^2}{\varphi c}
\nonumber \\ \approx 160
\frac{\alpha_B T_{{\rm w},2.8}^{1/2} A_{\star}^{1/2}
v_{{\rm w,}6}^{-1/2} t_{\rm d} v_{{\rm sh,}8.5}^2 } 
{\varphi_{1.3} (a_{14} - v_{{\rm sh,}8.5} t_{\rm d})}
~{\rm GeV}; ~\theta =0.
\label{E_max}
\ea
Here we assumed that $\varphi = 20\varphi_{1.3}$ is the number of
$e$-folding required for a thermal proton with $\sim 1$ keV kinetic
energy to reach $E_{p,\rm max}$.  For $\theta = \pi$, i.e. in the part
of the shell opposite to the RG, $E_{p,\rm max} \approx 90$ GeV at
$t=t_{\rm d}$~day.

Evolution of the shock velocity and radius in the deceleration phase
at later time ($\gtrsim 3$ day for $\theta \sim 0$), after the shell
accumulates swept up RG wind material of mass equal to its own in a
given solid angle, depends on the detail calculation for different
$\theta$ and is beyond the scope of this paper.  Here we simply assume
that protons can be accelerated to an energy up to the value in
Eq.~(\ref{E_max}) throughout the 15 day $\g$ ray emission episode, as
required by the $\pi^0$ model.

\section{Neutrino fluxes on Earth}
The observed $\g$ ray flux, averaged over 15 day outburst, from the
nova in V407 Cygni at a distance $D$ is fitted with a $\pi^0 \to 2\g$
decay model \cite{FermiV407Cyg} as
\ba
\Phi_{\gamma} (E_{\gamma}) \approx 
\frac{\langle n_H \rangle}{4\pi D^2} \epsilon_{\rm M} 
\int_{E_{p,th}}^{\infty} Q(E_p,E_\gamma) N_p(E_p) dE_p \,.
\label{gamma_flux}
\ea
Here $Q(E_p,E_\gamma)$ is the $\g$-ray production rate per unit
density of H atoms \cite{Kamae06}, $\epsilon_{\rm M} =$ 1.84 is the
nuclear enhancement factor to take into account the contribution of
other atoms \cite{Mori09},
$N_p(E_p)=N_{p,0}E_p^{-s}\exp[(m_pc^2-E_p)/E_{c,p}]$ is the
shock-accelerated proton spectrum and $\langle n_H \rangle$ is the
average number density of target particles.  The best-fit parameter
values are $s=2.15^{+0.45}_{-0.28}$, $E_{c,p} = 32^{+85}_{-8}$~GeV
with a total energy in protons $\approx 6.9^{+3.6}_{-2.3} \cdot
10^{42} (\langle n_H \rangle/4\cdot 10^8~{\rm cm}^{-3})^{-1}
(D/2.7~{\rm kpc})^2$~erg in steady state~\cite{FermiV407Cyg}.  Note
that a delta-function approximation \cite{Aharonian00} as $\Phi_\g
(E_\g)\propto (c\langle n_H \rangle /4\pi D^2 K_\pi) \int_{E_{\pi,\rm
th}}^\infty dE_\pi (E_\pi^2 - m_\pi^2 c^4)^{-1/2} \sigma_{pp} (m_p c^2
+ E_\pi/K_\pi) N_p (m_p c^2 + E_\pi/K_\pi)$ also produces reasonable
fit to the observed $\g$-ray flux with the same spectral parameters
$s$ and $E_{c,p}$.  Here $E_{\pi,\rm th} = E_\g + m_\pi^2c^4/4E_\g$ is
the threshold pion energy and $K_\pi \approx 0.17$ is the mean
fraction of the proton kinetic energy converting to $\pi^0$.

Neutrino fluxes from $\pi^\pm$ decays, which are created in the $pp$
interactions along with $\pi^0$, can be estimated from the observed
$\g$-ray source flux as~\cite{Gaisser,Lipari93,Costantini05}
\ba
\Phi^{\rm src}_{\nu_\mu} (E) = 
\Phi^{\rm src}_{{\bar \nu}_\mu} (E) = 0.50\Phi_\g (E); 
\nonumber \\ 
\Phi^{\rm src}_{\nu_e} (E) = 0.30\Phi_\g (E);~
\Phi^{\rm src}_{{\bar \nu}_e} (E) = 0.22\Phi_\g (E),
\label{nu_src_fluxes}
\ea
for $s=2.0$.  For softer indices, $s>2$, of the proton spectrum the
$\nu$ fluxes decrease compared to the $\g$-ray flux, following the
hadronic cascade theory~\cite{Gaisser}.  Contributions from kaons to
the neutrino fluxes are at the level of 10\% or less and we ignore
those.

\begin{figure}
\includegraphics[width=3.in]{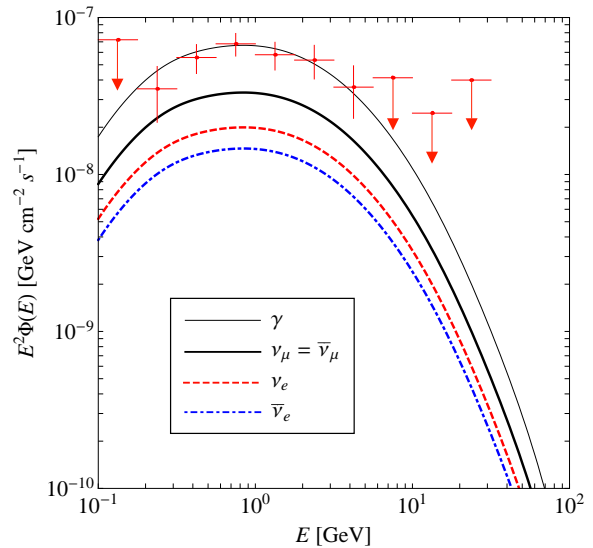}
\caption{Neutrino source fluxes (thick curves) from Nova 2010 in the
V407 Cygni symbiotic binary calculated, using
Eq.~(\ref{nu_src_fluxes}), from the best-fit $\pi^0$ decay model
parameters (thin solid curve) of $\g$ ray emission.  The data points
are taken from Ref.~\cite{FermiV407Cyg}}
\label{fig:nu_src_fluxes}
\end{figure}

Figure~\ref{fig:nu_src_fluxes} shows the $\nu$ fluxes (thick curves)
calculated using Eq.~(\ref{nu_src_fluxes}), and the $\pi^0$ decay
model (thin solid curve) for the $\g$ ray spectrum from Ref.\
\cite{FermiV407Cyg}.  The data points and 2-$\sigma$ upper limits are
also taken from Ref.\ \cite{FermiV407Cyg}.  Note that the source
fluxes will be modified as they are measured by any Earth-based
detector due to flavor oscillation in vacuum and in matter (MSW
effect) inside the Earth, which is particularly important below $\sim
50$ GeV.

High-energy $\nu$'s, created with definite flavors ($\nu_\alpha$ with
$\alpha = e,\,\mu,\,\tau$), arrive as coherent mass eigenstates
($\nu_i$ with $i=1,\,2,\,3$) from V407 Cygni on the Earth's surface.
This is because the separation between the wave packets of different
mass eigenstates $d_s = L\Delta m^2/2E_\nu^2 \approx 2.4\cdot 10^{-4}
(L/{2.7\,\rm kpc}) (\Delta m_{31}^2/2.4\cdot 10^{-3}\,{\rm
eV}^2)(E_\nu/{10\,\rm GeV})^{-2}$~cm, arising from propagation, is
much smaller than the typical size of the $\nu$ wave packets from
$\pi^\pm$ and $\mu^\pm$ decays.  However, the $\nu$ detectors do not
have necessary energy resolution, $\Delta E/E\approx 10^{-13}
(L/{2.7\,\rm kpc})^{-1} (E_\nu/10\,{\rm GeV})$, to measure the
oscillatory pattern arising from coherent $\nu$ mass eigenstates at
the detectors, and can only measure averaged oscillation.  This loss
of coherence by the detector during measurement is equivalent to
incoherent $\nu$ mass eigenstates arriving at the surface of the Earth
(see, e.g. Ref. \cite{HEnuOsc}).  The resulting total $\nu$ flavor
conversion probability at a detector is
\ba
P_{\nu_\alpha\to\nu_\beta} &=& 
\sum_i  P^{\rm src}_{\nu_\alpha\to\nu_i} 
P^{\oplus}_{\nu_i\to\nu_\beta} \nonumber \\ 
&=& \sum_i |U_{\alpha i}|^2 | \sum_\eta 
A^{\oplus}_{\beta\eta} U_{\eta i} |^2 .
\label{total_prob}
\ea
Here $P^{\rm src}$ and $P^\oplus$ are conversion probabilities from
the flavor to mass and from the mass to flavor states, respectively.
The PMNS mixing matrix for the flavors and mass eigenstates are
denoted with $U$, and $A^{\oplus}$ is the transition amplitude after
propagation inside the Earth.  The conversion probability at the
source is neglected in the second equality, which is valid because of
a negligible matter potential at the $\nu$ production site.  (However
an interesting situation may arise if $\nu$'s pass through the RG,
which will modify the source probablity.)  \footnote{For $\gtrsim 50$
GeV $\nu$'s the Earth's matter potential is large enough to suppress
oscillation and the probability simplifies as
$P^\oplus_{\nu_i\to\nu_\beta} = |U_{i\beta}|^2$.}

\begin{figure}
\includegraphics[width=3.5in]{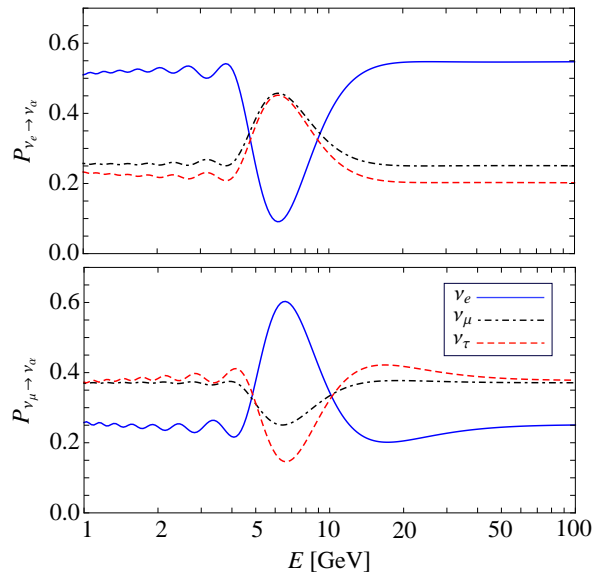}
\caption{Flavor conversion probabilities [Eq.~(\ref{total_prob})] for
neutrinos created at V407 Cygni and detected by an Earth-based
detector at the South pole.  The $\nu_e \to \nu_\alpha$ and $\nu_\mu
\to \nu_\alpha$ probabilities are plotted in the upper and lower
panel, respectively.  We used normal hierarchy of $\nu$ masses and a
vanishing CP phase along with the best-fit oscillation paramaters (see
text for details).}
\label{fig:nu_osc_prob} 
\end{figure}

For the position of V407 Cygni, Right Ascension $(RA)=315.55^\circ$
and Declination $(DEC)=45.74^\circ$, $\nu$'s travel a path length
$L=2R_\oplus\cos\theta_n$ inside the Earth, with radius $R_\oplus$,
and a nadir angle $\theta_n = 90^\circ - DEC$.  We calculate
$A^{\oplus}$ with a numerical code \cite{MMR08} that uses the
Preliminary Reference Earth Model \cite{PREM} for the density profile
inside the Earth, and $\nu$ mixing parameters: $\Delta m_{31}^2 =
2.4\cdot 10^{-3}$ eV$^2$, $\Delta m_{21}^2 = 8\cdot 10^{-5}$ eV$^2$,
$\sin^2\theta_{12} = 0.31$, $\theta_{23} = \pi/4$, $\sin^2\theta_{13}
= 0.02$ and the CP violating phase $\delta = 0$.  We consider normal
$\nu$ mass hierarchy only.  The total conversion probabilities are
plotted in Fig.~\ref{fig:nu_osc_prob} for neutrinos.  The upper and
lower panels correspond to $\nu_e \to \nu_\alpha$ and $\nu_\mu \to
\nu_\alpha$ conversions, respectively.  The anti-neutrino conversion
probablities are not affected by matter in case of normal $\nu$ mass
hierarchy and vacuum conversion formalism apply.

The most prominent features in the probability curves, the dip in the
$P_{\nu_e \to \nu_e}$ (and associated peaks/dips for other $\nu$'s)
at $\sim 6$ GeV, can be understood analytically from 2-flavor $\nu$
oscillation framework with 1-3 mixing \cite{Akhmedov07}.  For
$\theta_n \gtrsim 34^\circ$, which is the case for V407 Cygni, $\nu$'s
do not pass through the Earth's core.  Conversions mostly take place
in the mantle with an average density of $\langle\rho\rangle\sim 5$
g~cm$^{-3}$.  For this density, the low MSW resonance energy is $E_L =
\Delta m_{12}^2\cos 2\theta_{12}/[2\sqrt{2} G_{\rm F}
\langle\rho\rangle] \approx 0.08$ GeV and 1-2 mixing is strongly
suppressed for our energy range of interest, $E_\nu \gtrsim 1$ GeV.
The dip at $\sim 6$ GeV for $P_{\nu_e \to \nu_e}$ in
Fig.~\ref{fig:nu_osc_prob} ({\em upper panel}) corresponds to the 1-3
or high MSW resonance energy $E_H = \Delta m_{13}^2\cos
2\theta_{13}/[2\sqrt{2} G_{\rm F} \langle\rho\rangle] \approx 6$ GeV.
The width of the dip is $2\tan 2\theta_{13}E_H \approx 3.6$ GeV.  At
energies $\gg E_H$, the conversion probablities are dominated by
vacuum oscillation.

The $\nu$ fluxes at a detector buried under the Earth's surface are
calculated from Eqs.~(\ref{nu_src_fluxes}) and (\ref{total_prob}) as
\ba
\Phi_{\nu_\alpha}^{\rm det} = \Phi_{\nu_\mu}^{\rm src} 
P_{\nu_\mu \to \nu_\alpha} + \Phi_{\nu_e}^{\rm src} 
P_{\nu_e \to \nu_\alpha}\,.
\label{nu_det_fluxes}
\ea
Event rates in a $\nu$ detector depend on these fluxes and we
calculate the expected number of events from a $\g$-ray nova outburst
such as Nova 2010 in V407 Cygni next.

\section{High Energy Neutrino Detection}
The total number of neutrino-nucleon charge current (cc) interactions
by $\nu_\alpha$, which produce secondary leptons of type $\alpha$ that
may be detectable, can be calculated as
\ba
N_{\nu_\alpha} = \frac{N_{\rm T}t}{V_{\rm det}}
\int dE_\nu 
\int d\Omega  ~V_{\rm eff} (E_\nu, \Omega)
\nonumber \\ 
\times \left[ 
\sigma_{\nu}^{\rm cc} (E_\nu) 
\Phi^{\rm det}_{\nu_\alpha} (E_\nu, \Omega) +
\sigma_{\bar \nu}^{\rm cc} (E_\nu) 
\Phi^{\rm det}_{{\bar \nu}_\alpha} (E_\nu, \Omega)
\right],
\label{event_count}
\ea
below the energy ($\sim 100$ TeV) at which $\nu$ absorption inside the
Earth is negligible (see e.g. \cite{MMR08}).  Here $t$ is the
duration of the $\nu$ outburst, $V_{\rm det}$ is the instrumented
detector volume and $N_{\rm T}$ is the total number of nucleons in
that volume.  The effective detector volume $V_{\rm eff} \approx
V_{\rm det}$ for contained events at low energies.  At very high
energies $V_{\rm eff} > V_{\rm det}$, specially for $\nu_\mu$ because
secondary $\mu$'s that are produced far from the physical detector can
propagate inside due to their long range.  We parameterize the cc
cross-sections in the $\sim 10$ - 100 GeV range \cite{Gandhi98} as
\ba
\sigma^{\rm cc}_\nu = 7.3\cdot 10^{-39} (E_\nu/{\rm GeV})~{\rm cm}^2
\nonumber \\
\sigma^{\rm cc}_{\bar \nu} = 
3.8\cdot 10^{-39} (E_\nu/{\rm GeV})~{\rm cm}^2 .
\label{cc_cross_section}
\ea

For a detector such as the IceCube Deep Core \cite{DeepCore}, with a
10 Mt fiducial volume, the total number of nucleons inside $V_{\rm
det}$ is $N_{\rm T} = 10^{13}N_A$, where $N_A$ is the Avogadro's
number.  All astrophysical sources are essentially point sources in
$\nu$'s and one needs to carry out integration over $E_\nu$ only in
Eq.~(\ref{event_count}).  The lower range $E_{\nu,\rm th}$ depends on
the energy of the secondary lepton at the detection threshold.  The
average muon energies are $\approx 0.52E_\nu$ for neutrinos and
$\approx 0.66E_\nu$ for anti-neutrinos in the range $E_\nu \sim 10$ -
100 GeV.  For a rough estimate of $N_{\nu_\alpha}$ we assume a common
$E_{\nu,\rm th}$ for both $\nu$'s and ${\bar \nu}$'s for the same
lepton energy.  Here we consider $\nu_\mu$'s and ${\bar \nu}_\mu$'s
only, as the detection prospect of $\sim 5$ GeV muon tracks is better
than the showers created by other flavors.  For $t=15$ day, the
duration of the $\g$ ray outburst from the nova in V407 Cygni, and for
$V_{\rm eff} = V_{\rm det}$ we calculate
$$
N_{\nu_\mu} \approx 0.5^{+4.4}_{-0.4} ~, 
~ N_{{\bar \nu}_\mu} \approx 0.3^{+2.3}_{-0.2} ~;~~
10 \le E_\nu/{\rm GeV} \le 100,
$$
using the $\pi^0$ model for $\g$ rays with the best-fit parameters,
and their $\pm 1\sigma$ variation.

\section{Discussion and implications} We have calculated $\sim
0.1$-100~GeV $\nu$ fluxes from the $\g$ ray Nova 2010 in the symbiotic
binary V407 Cygni, from $\pi^\pm$ decays while assuming that the $\g$
rays detected by the {\em Fermi} LAT are from $\pi^0$ decays.  
Detection of these $\nu$'s in the $\sim 10$-100~GeV range depends
critically on the detector's angular resolution and uncertainty in the
atmospheric $\nu$ background.  For the IceCube Deep Core the
atmospheric $\nu_\mu + {\bar \nu}_\mu$ events during 15 days of the
$\g$-ray nova is $\sim 60$ for $\Delta\theta = 10^\circ$ following
Ref.\ \cite{MMR08}.  Future large scale detectors with a better
angular resolution, such as $\sim 3^\circ$ in the
Super-Kamiokande~\cite{sk1}, can reduce the background roughly by an
order of magnitude to improve the signal-to-noise ratio, as well as
measure background more precisely to reduce uncertainty.  Moreover, a
similar $\g$-ray nova in the nearest symbiotic nova system AG Peg
($D\sim 0.6$~kpc), among a dozen known, would improve the signal by a
factor of $\sim 20$.  More energetic $\g$ ray nova is also possible.
The kinetic energy release in 2006 nova outburst in a nearby ($D\sim
1.4$ kpc) symbiotic system RS Ophiuchi is inferred to be an order of
magnitude higher~\cite{laming09} than in V407 Cygni 2010
nova~\cite{FermiV407Cyg}.

The rate of classical novae in our galaxy is $\sim 30$ year$^{-1}$ in
all binary systems \cite{nova_rate}.  The rate of novae among
$\sim 100$ known symbiotic binary systems~\cite{symbiotic}, which seem
to require for particle acceleration and $\g$ ray production, is more
uncertain, once every few years.  Following up of future novae with
$\nu$ detectors may establish a rate independent of the $\g$ ray
instruments and answer the question of a hadronic or leptonic origin
of the $\g$ rays from them.  Non-detection of $\nu$'s from $\g$ ray
novae can be used to constrain the emission models and system
parameters as well.

While astrophysical $\g$ ray sources have been hypothesized as
potential TeV - PeV $\nu$ sources, mostly from $\pi$'s and $K$'s
created in $p\g$ interactions, relatively low-energy ($\sim 10$ GeV)
$\nu$ sources which are simultaneous $\g$ ray sources were thought to
be rare.  Surprising discovery of the nova in V407 Cygni by the {\em
Fermi} LAT adds a new $\nu$ source candidate class where, like
supernova remnants, $pp$ interactions may produce both $\g$ rays and
$\nu$'s.

\par

We thank C.C.~Cheung, C.D.~Dermer, A.Yu.~Smirnov, C.~Spiering and
K.S.~Wood for helpful comments and discussion.  Work of SR is
supported by NASA Fermi Cycle II Guest Investigator Program and the
National Research Council.

\end{document}